\documentclass[runningheads]{llncs}
\usepackage{graphicx}
\usepackage{cite}
\usepackage{amsmath,amssymb,amsfonts}
\usepackage{algorithmic}
\usepackage{graphicx}
\usepackage{textcomp}
\usepackage{xcolor}
\usepackage{float}
\usepackage{url}
\usepackage{csquotes}
\usepackage{comment}

\usepackage[capitalize]{cleveref}

\usepackage{tikz}
\usetikzlibrary{shapes.multipart, shapes, positioning, trees}
\tikzstyle{boxnode}=[draw, fill=gray!10]

\usepackage[nolist]{acronym}
\def\BibTeX{{\rm B\kern-.05em{\sc i\kern-.025em b}\kern-.08em
    T\kern-.1667em\lower.7ex\hbox{E}\kern-.125emX}}

\begin{document}
\title{IT/OT Integration by Design
\thanks{This preprint has not undergone peer review or any post-submission
improvements or corrections.}}
%
%
\author{Georg Schäfer\inst{1} \and
Hannes Waclawek\inst{1} \and
Sarah Riedmann\inst{2} \and
Christoph Binder\inst{2} \and
Christian Neureiter\inst{2} \and
Stefan Huber\inst{1}}
\authorrunning{G. Schäfer et al.}
%
\institute{Josef Ressel Centre for Intelligent and Secure Industrial Automation, Austria \and
Josef Ressel Centre for Dependable System-of-Systems Engineering, Austria \\
\email{\{firstname.lastname\}@fh-salzburg.ac.at}
}
\maketitle              
\begin{abstract}
The four Industry 4.0 design principles information transparency, technical assistance, interconnection, and decentralized decisions pose challenges in integrating \ac{IT} and \ac{OT} solutions in industrial systems. These different solutions have conflicting requirements, making interfaces between them problematic for both systems and organizations.
An \ac{IBPT} entity, acting as an intermediary between the realms of \ac{IT} and \ac{OT}, has been proposed in a previous work, to effectively reduce the amount of required \ac{IT}/\ac{OT} interfaces in an attempt of overcoming this situation.
In this work, we investigate the effects of this approach during the design phase. We argue that, by eliminating interfaces between \ac{IT} and \ac{OT} components in the system design, this approach is therefore eliminating conflicting communication channels within the organization's communication structure. In order to verify our argument, we develop a model of our 
\ac{IBPT} concept according to the \ac{RAMI} using an Industry~4.0 scenario addressing the four essential Industry~4.0 design principles. Results show that the \ac{IBPT} approach indeed eliminates potentially conflicting \ac{IT}/\ac{OT} interfaces during the system design phase.

\end{abstract}

\begin{acronym}[SQL]
 \acro{AR}[AR]{Augmented Reality}
 \acro{ATAM}[ATAM]{Architecture Tradeoff Analysis Method}
 \acro{RAMI}[RAMI4.0]{Reference Architecture Model Industrie 4.0}
 \acro{IBPT}[IBPT]{Industrial Business Process Twin}
 \acro{IoT}[IoT]{Internet-of-Things}
 \acro{IT}[IT]{information technology}
 \acro{OT}[OT]{operational technology}
 \acro{IIoT}[IIoT]{Industrial Internet of Things}
 \acro{MDG}[MDG]{model-driven generation}
 \acro{EV}[EV]{Electric Vehicle}
 \acro{DT}[DT]{Digital Twin}
 \acro{XMI}[XMI]{XML Metadata Interchange}
 \acro{BMS}[BMS]{Battery Management System}
 \acro{HLUC}[HLUC]{High Level Use Case}
 \acro{BUC}[BUC]{Business Use Case}
 \acro{PUC}[PUC]{Primary Use Case}
 \acro{PoC}[PoC]{Proof-of-Concept}
 \acro{AD}[AD]{Application Domain}
 \acro{SoS}[SoS]{System-of-Systems}
 \acrodefplural{SoS}[SoS]{Systems-of-Systems}
 \acro{ADSRM}[ADSRM]{Agile Design Science Research Methodology}
 \acro{DSR}[DSR]{Design Science Research}
 \acro{CIM}[CIM]{Computation Independent Model}
 \acro{PIM}[PIM]{Platform Independent Model}
 \acro{PSM}[PSM]{Platform Specific Model}
 \acro{MBSE}[MBSE]{Model-based Systems Engineering}
 \acro{MDA}[MDA]{Model Driven Architecture}
 \acro{DSL}[DSL]{Domain-Specific Language}
 \acro{GPL}[GPL]{General Purpose Language}
 \acro{DSSE}[DSSE]{Domain Specific Systems Engineering}
 \acro{SPES}[SPES]{Software Platform Embedded Systems}
 \acro{SGAM}[SGAM]{Smart Grid Architecture Model}
 \acro{ARAM}[ARAM]{Automotive Reference Architecture Model}
 \acro{IBPT}[IBPT]{Industrial Business Process Twin}
 \acro{UML}[UML]{Unified Modeling Language}
 \acro{CPS}[CPS]{Cyber-physical System}
 \acrodefplural{CPS}[CPS]{Cyber-physical Systems}
 \acro{SOA}[SOA]{Service-oriented Architecture}
 \acro{ADL}[ADL]{Architecture Description Language}
 \acro{FAS}[FAS]{Functional Architecture for Systems}
 \acro{HMI}[HMI]{Human-Machine Interface}
 \acro{MDE}[MDE]{Model-driven Engineering}
 \acro{EA}[EA]{Enterprise Architect}
 \acro{AI}[AI]{Artificial Intelligence}
 \acro{SoI}[SoI]{System of Interest}
 \acro{SuD}[SuD]{System under Development}
 \acro{SPES}[SPES]{Software Platform Embedded Systems}
 \acro{IIRA}[IIRA]{Industrial Internet Reference Architecture}
 \acro{UAF}[UAF]{Unified Architecture Framework}
 \acro{SysML}[SysML]{Systems Modeling Language}
 \acro{ICT}[ICT]{Information and Communication Technology}
 \acro{UAF}[UAF]{Unified Architecture Framework}
 \acro{AAS}[AAS]{Asset Administration Shell}
 \acro{API}[API]{Application Programming Interface}
 \acro{UAF}[UAF]{Unified Architecture Framework}
 \acro{RTE}[RTE]{Round-trip Engineering}
 \acro{JSON}[JSON]{JavaScript Object Notation}
 \acro{BPMN}[BPMN]{Business Process Model and Notation} 
 \acro{NFC}[NFC]{Near-field Communication} 
 \acro{XML}[XML]{Extensible Markup Language}
 \acro{OMG}[OMG]{Object Management Group}
 \acro{ISDSR}[IS DSR]{Information Systems Design Science Research}
 \acro{CAEX}[CAEX]{Computer Aided Engineering Exchange}
 \acro{GUI}[GUI]{Graphical User Interface} 
 \acro{SME}[SME]{Small and Medium-sized Enterprise}
 \acro{DLL}[DLL]{Dynamic-Link Library}
 \acro{PLC}[PLC]{Programmable Logic Controller}
 \acro{NLP}[NLP]{Natural Language Processing}
 \acro{CV}[CV]{Computer Vision}
 \acro{OPCUA}[OPC~UA]{Open Plattform Communication Unified Architecture}
\end{acronym}

\section{Introduction}\label{sec:int}

The integration of \ac{IT} solutions, commonly encountered at the office floor level of a production facility, with \ac{OT} solutions present on the shop floor, frequently poses a significant challenge. 
Essentially different characteristics of the worlds of \ac{IT} and \ac{OT}, as described in~\cite{stouffer2015}, bring together different stakeholders with often contradicting or conflicting requirements. 
This leads to conflicting communication channels in the organization.
Conway's law, originally introduced in~\cite{conway1968committees} and later supported by evidence under the term \enquote{mirroring hypotheses} in~\cite{macormack2012}, illustrates how these conflicting communication channels are reflected in the system design, as shown on the left-hand side of \cref{FIG_Conway}. 
In this sense, conflicts between \ac{IT} and \ac{OT} stakeholders within the organizational structure are also present in the \ac{IT}/\ac{OT} system design.
To ensure the interconnection of all system components and to enable early verification or validation, system architecture models are core elements for describing the system before actually implementing it~\cite{sahu2020review}. Applying this to the \ac{IT}/\ac{OT} integration task, creating system architectures for \ac{IT} and \ac{OT} systems can contribute to finding necessary interfaces between both worlds~\cite{lipnicki2018future}. This is beneficial for bringing together stakeholder perspectives and finding a common language and mutual base for discussions. \ac{RAMI} as described in \cite{hankel2015reference} has been proposed to address challenges for the design of Industry~4.0 production systems. \ac{MBSE} has become the main driver for developing systems that are based on \ac{RAMI}, however, it is often difficult to apply the theoretical concepts of \ac{RAMI} across all of its views and concerns. The RAMI Toolbox proposed in~\cite{binder2021towards} is an approach to overcoming these difficulties. 

\subsection{Contribution}

We build upon our previous work documented in~\cite{waclawek2023}, where we introduce the \ac{IBPT} concept with an \ac{OPCUA} communication and information model, enabling \ac{IT}/\ac{OT} integration. It decouples the realms of \ac{IT} and \ac{OT} by acting as an intermediary between both worlds, thus allowing for an integration of \ac{IT} and \ac{OT} components of different manufacturers and platforms.
In this work, we investigate the effects of this approach during the design phase. We argue that, by eliminating interfaces between \ac{IT} and \ac{OT} components in the system design, this approach is therefore eliminating conflicting communication channels within the organization's communication structure. 
In order to evaluate our argument, we utilize the \ac{RAMI} Toolbox to develop a system model for the Nine Men's Morris case study introduced in~\cite{waclawek2023}. This scenario consists of two geographically distributed robot cells playing the two-player board game, demonstrating the four essential Industry~4.0 design principles. The resulting \ac{EA} model is provided at~\cite{schaefer2023}.

\subsection{Related work}
With the goal to fully exploit the potential of Industry~4.0, many companies strive to interconnect their shop floor systems with their office floor systems~\cite{hicking2021collaboration}. \ac{IT} or enterprise system concepts integrate or extend business processes across the boundaries of the enterprise's functions, either within the organization or between organizations~\cite{da2011enterprise}. \ac{OT} production system concepts are used to monitor, control and execute production tasks. Due to the emergence of data-driven methods that come along with the essential Industry~4.0 design principles outlined in~\cite{hermann2016design}, these two system concepts are in need of approaching each other. While resulting in new opportunities for manufacturing companies, new challenges, like cybersecurity, arise~\cite{murray2017convergence}. Thus, in order to push forward the digital transformation for those companies, novel use cases are approached. Accompanied by a set of requirements, typical challenges that come along with \ac{IT}/\ac{OT} integration are elaborated in~\cite{hicking2021collaboration}.

In \cite{hicking2021}, Hicking et al. present an approach for integrating \ac{IT} and \ac{OT} systems by selecting relevant digitalization use cases of an organization. The current \ac{IT}/\ac{OT} landscape of an organization is assessed using a unified use case for \ac{IT}, \ac{OT} and interface components. This profile is then matched against the requirements of utility use cases to identify integration efforts. The paper outlines some example use cases and describes the components of the proposed IT/OT integration profile. 
In this work, we put our focus not on Brownfield, but Greenfield systems at design time.
The authors of \cite{piotr2018} discuss the problem of \ac{IT}/\ac{OT} integration in the context of oil and gas production. Besides defining recommended security levels for production organizations in these fields, they argue that a key to successful \ac{IT}/\ac{OT} integration is the effective collaboration between departments at the boundary between manufacturing and enterprise zones of an organization. This is in accordance with our approach of eliminating potentially conflicting \ac{IT}/\ac{OT} communication channels within the organization, to make processes at the boundary between both worlds more streamlined and effective.
In~\cite{palade2021open}, an approach for an open platform for smart production addressing \ac{IT}/\ac{OT} integration within a factory is proposed.
Production planning is realized by an assisted production control demonstrator, while the hybrid architecture designing demonstrator addresses the need for vertical integration of \ac{IT} systems based on a system architecture. Contrary to this work, we do not introduce a new framework, but rather build upon \ac{RAMI} and extend it to ensure the \ac{IT}/\ac{OT} integration on a system-level perspective.


\section{Approach} \label{sec:approach}

\begin{figure}[tb]
\begin{center}
	\includegraphics[width=1.0\textwidth]{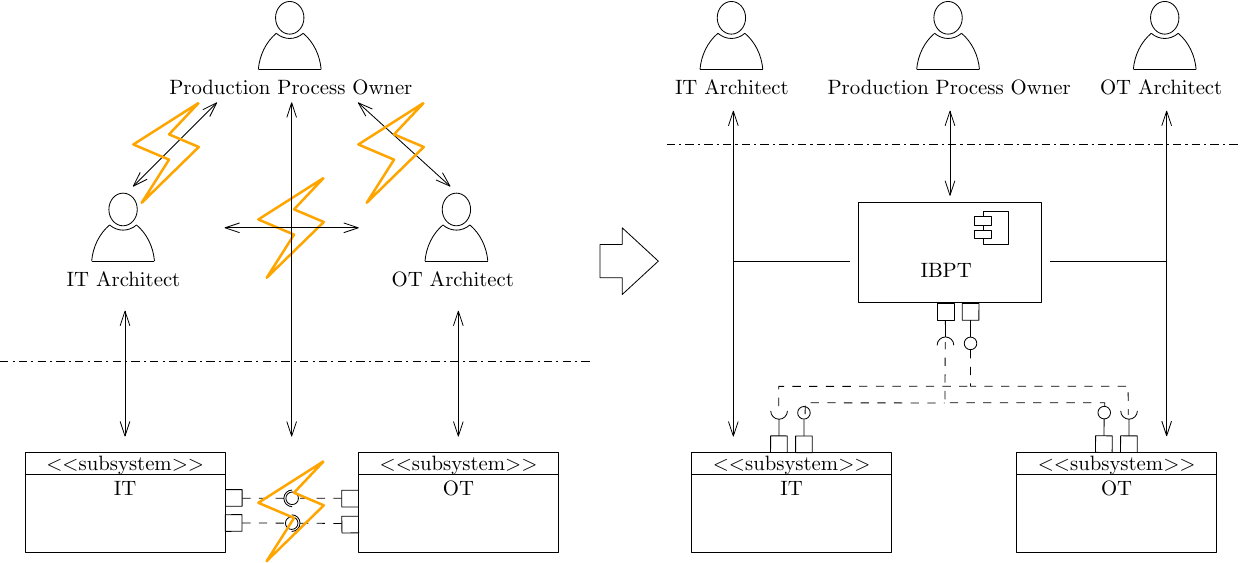}
\end{center}
	\caption{By Conway's law, communication structures within the organisation are represented in the system's design. Our approach eliminates interfaces between \ac{IT} and \ac{OT} components in the system design, which, by the mirroring hypotheses, therefore eliminates conflicting communication channels also within the organization's communication structure.}
	\label{FIG_Conway}
\end{figure}

By Conway's law, communication structures within the organisation are represented in the system's design and vice versa. Potentially conflicting communication channels present in classical organization structures required for production processes between process owners and architects for IT and OT systems therefore transfer to the IT/OT system design. By introducing an \ac{IBPT} entity as a workflow engine acting as an intermediary between IT and OT components, we eliminate interfaces between \ac{IT} and \ac{OT} components in the system design, which, by the mirroring hypotheses, therefore eliminates conflicting communication channels also within the organization's communication structure.
This principle is depicted in \cref{FIG_Conway}. The mirror axis is indicated as dash dotted line.

\subsection{Eliminating IT/OT communication structures}

The left-hand side of \cref{FIG_Conway} shows the situation at design time without \ac{IBPT}. 
For a new production scenario, requirements and design constraints need to be negotiated between the process owner and IT as well as OT architects. As per Conway's law, the system design then reflects 
this via interfaces between IT and OT components. 
By the \ac{IBPT} concept on the right-hand side of \cref{FIG_Conway}, the process owner defines  
interfaces, an information model as well as workflows using \ac{BPMN}. Data types and interfaces are abstracted to the level of business logic. These definitions are then used by IT and OT architects to realize necessary interfaces within their respective components. In this way, the common language of
involved stakeholders is abstracted to the level of business logic, thus making it easier to understand and meet requirements for both sides. At runtime, in the sense of \ac{BPMN}, an instance of the \ac{IBPT} entity acts as a workflow engine. Every interaction with the engine spawns a new business process, that is in bidirectional exchange with IT and OT components, twinning the state of the business process. This principle is shown in \cref{FIG_IBPT}.

\subsection{Industrial Business Process Twin (IBPT)} \label{sec:ibpt}
Kritzinger et al. proposed a taxonomy for \acp{DT} and related mirroring entities, based on the nature of data flow between the object and the mirroring entity~\cite{kritzinger2018digital}.
When data flow is carried out manually in both directions, a Digital Model is employed. Conversely, if the data flow is manual in one direction and automatic in the other, this constitutes a Digital Shadow.
Only if the flow of data is automated in both directions can the system be classified as a \ac{DT}. 
Kritzinger et al. also argue that a \ac{DT} is not restricted to mirror physical objects.
The replication of virtual entities, like business processes, can be realized using \acp{DT} as well, as long as the communication in both directions takes place automatically.


  
\begin{figure}[tb]
\begin{center}
	\includegraphics[width=1.0\textwidth]{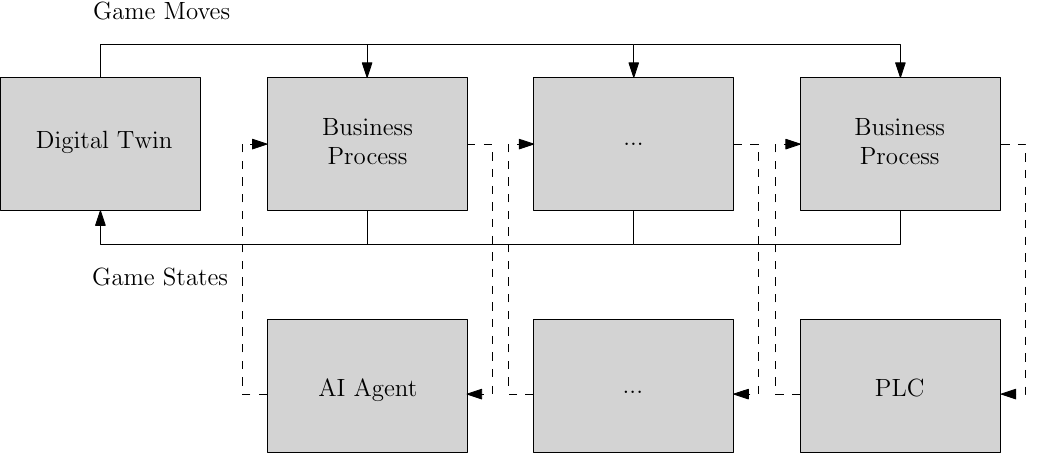}
\end{center}
	\caption{Simplified illustration of the \ac{IBPT} concept of twinning industrial business processes. The \ac{DT} is in bidirectional exchange with \ac{OT} and \ac{IT} components. Interactions with business processes are represented by solid lines, whereas interactions with \ac{IT}/\ac{OT} components are represented by dashed lines. It thus decouples \ac{OT} and \ac{IT} concerns by acting as an intermediary / orchestrator between both worlds. Every interaction spawns a new process that orchestrates activities carried out by \ac{IT} and \ac{OT} components. In the sense of \ac{BPMN}, this follows the principle of a workflow engine.}
	\label{FIG_IBPT}
\end{figure}

Based on this principle of twinning business processes, we introduce the \ac{IBPT} concept in~\cite{waclawek2023}.
An \ac{IBPT} entity is placed at the intersection of the worlds of \ac{IT} and \ac{OT},
between the Enterprise and Workstation layers of the \ac{RAMI} Hierarchical perspective.
From an \ac{IT} perspective, this leaves us
with centering our modeling approach around business
processes at the highest level, since business processes
reside on the Enterprise segment of the Business layer.
Since a core idea of the \ac{SOA} paradigm is centering around
business activities, this enables a \ac{SOA}-centric architecture design~\cite{openGroup2009}.
The functionality of the \ac{OT} subsystem is abstracted throughout the layers of the \ac{RAMI}
Architectural perspective, providing unified interfaces for interacting with the \ac{OT} subsystem.
In this way, an \ac{IBPT} entity is acting as an intermediary between the worlds of \ac{IT} and \ac{OT} in an attempt to reduce \ac{OT} system complexity for \ac{IT} stakeholders and vice versa. 
Abstracted system functionality is modeled using a semantic information modeling approach, holistically spanning the layers of \ac{RAMI}. 
Considering our scenario of playing the game Nine Men's Morris as described in \cref{sec:scenario}, in the sense of \ac{BPMN}, it acts as a workflow engine for production processes. A production order spawns production process instances, which are in bidirectional exchange with \ac{IT} and \ac{OT} components. This principle is depicted in \cref{FIG_IBPT}. Interactions with business processes are represented by solid lines, whereas interactions with \ac{IT}/\ac{OT} components are represented by dashed lines.

\subsection{Industry~4.0 scenario}\label{sec:scenario}

\begin{figure}[tb]
\begin{center}
	\includegraphics[width=1.0\textwidth]{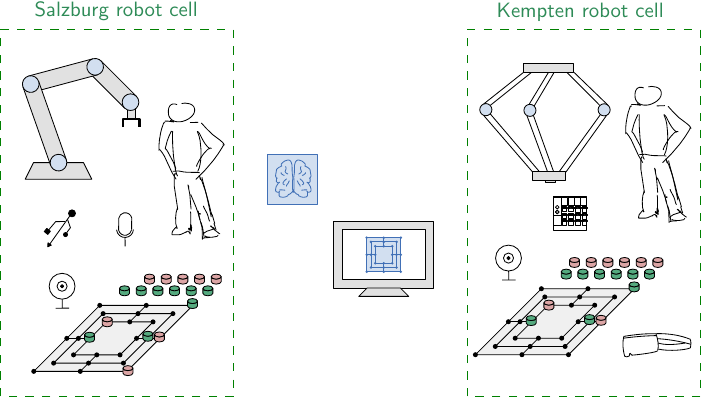}
\end{center}
	\caption{Components used in the scenario of playing the game of Nine Men's Morris, demonstrating the four essential Industry~4.0 design principles introduced in~\cite{hermann2016design}.}
	\label{FIG_SCENARIO}
\end{figure}

The utilization of game-based scenarios is beneficial for the use in academic research and education. For one, literature shows that gamification is beneficial for education \cite{nah2014}. For another, game-based scenarios provide well-defined environments, states, and actions for training AI agents. In this context, Nine Men’s Morris is a well-suited example for a pick-and-place production task and AI-based optimization. The shape of game tokens used in the game is favorable for different kinds of robot grippers. The game mechanics are characterized by their simplicity, which is why they can be learned quickly. This is why the Salzburg University of Applied Sciences together with the Kempten University of Applied Sciences center one of their testbeds for AI-related research and education around geographically distributed robot cells playing the game of Nine Men’s Morris. In this work, we want to harness this infrastructure and develop an \ac{IBPT}-centric system architecture that integrates both IT and OT components using \ac{MBSE} via the RAMI Toolbox described in \cref{sec:ramitoolbox}. A single robot cell consists of the game board with its tokens, various \ac{HMI} devices like a \ac{NLP} microphone, a \ac{CV} camera, \ac{AR} glasses and a robot performing the pick-and-place task of moving game tokens across the game board. To demonstrate the integration of hardware of different manufacturers and platforms, one cell utilizes a delta robot controlled by a \ac{PLC} and the other one utilizes an USB-controlled robot arm. Game moves are sent to the robots via a web-based \ac{GUI}, an \ac{AI} agent, \ac{NLP}, \ac{CV} or \ac{AR} devices within the robot cell in question. The concept is depicted in \cref{FIG_SCENARIO}, illustrating the aforementioned components with their geographical location.

In the following, we argue that this scenario is in accordance with the four essential Industry~4.0 design principles introduced by Hermann et al. in~\cite{hermann2016design}. Information transparency brings the need for a holistic semantic information modeling approach.
In this way, data aggregation and analysis can be performed by different stakeholders via well-defined interfaces with a common and mutual understanding of the data in question.
In our scenario, we establish this design principle by utilizing a holistic semantic \ac{OPCUA} information model.
In order to enable interconnection, again, a holistic,
multi-factory approach considering geographic distribution
requires the use of common communication standards
allowing information exchange of different systems over the internet. 
This design principle is demonstrated through \ac{OPCUA}-based communication between geographically distributed robot cells.
The principle of decentralized decisions is incorporated by geographically distributed players and \ac{AI} players that control the game flow remotely.
Technical assistance benefits from implementing modern \acp{HMI} allowing human-robot
interaction in an intuitive way that is beneficial
for the use in industrial environments. 
This is realized in our scenario by incorporating different \acp{HMI} like \ac{CV}, \ac{NLP} or \ac{AR} for human-machine interaction.
Following this line of argumentation, our scenario meets the essential Industry~4.0 design principles.
A detailed overview of the described scenario is presented in~\cite{waclawek2023}.

\subsection{RAMI Toolbox} \label{sec:ramitoolbox}
\begin{figure}[tb]
\begin{center}
	\includegraphics[width=1.0\textwidth]{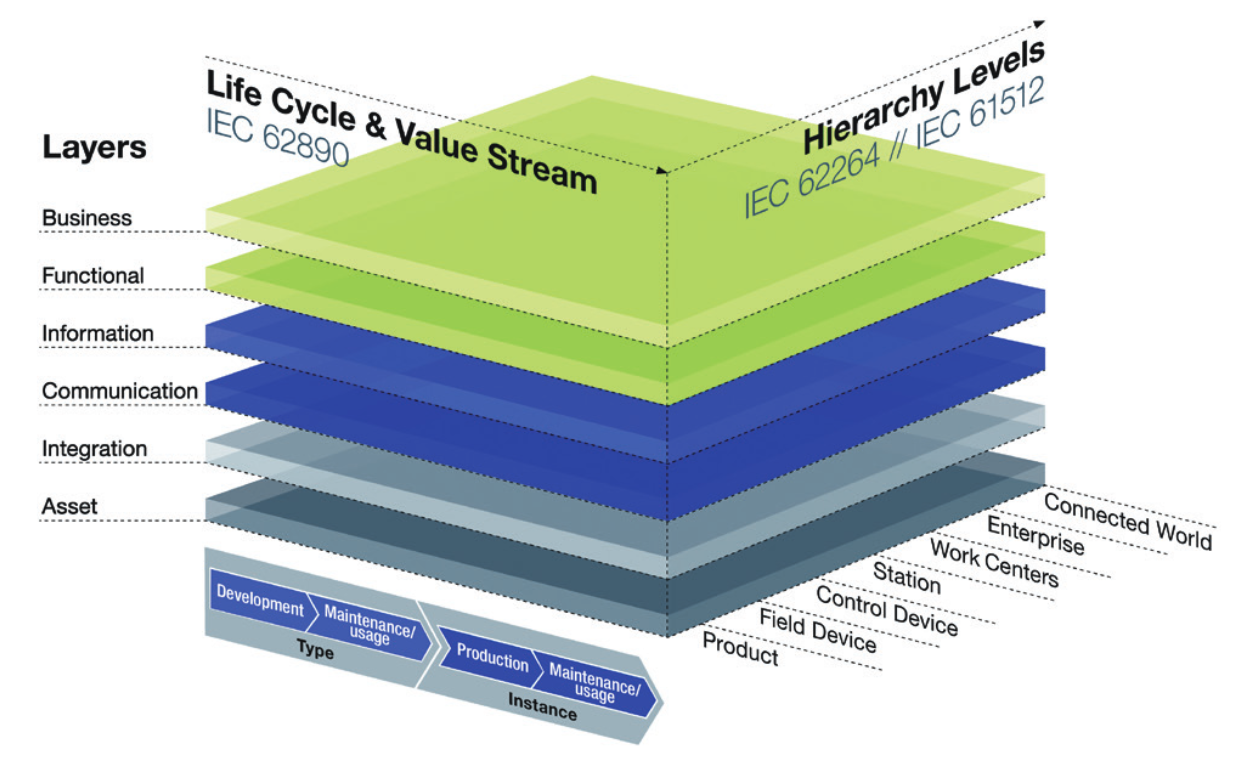}
\end{center}
	\caption{Reference Architecture Model Industrie 4.0 (RAMI 4.0) as shown in \cite{bitkom2015strategy}.}
	\label{FIG_RAMI}
\end{figure}

\ac{RAMI} was introduced to establish a common basis for standardization and mutual understanding in the domain of Industry~4.0~\cite{hankel2015reference}. This led to its publication as a German DIN SPEC 91345 standard~\cite{spec201691345}. The \ac{RAMI} model, illustrated in \cref{FIG_RAMI}, comprises three axes: the \enquote{Interoperability Layers}, the horizontal \enquote{Life Cycle \& Value Stream}, as well as ``Hierarchy Levels'' axis, each addressing distinct aspects of an industrial system. The top-down arranged layers offer different viewpoints of the industrial system. The ``Hierarchy Levels'' axis embeds an extended version of the ISA-95 automation pyramid by considering the system at different scales, from a connected world to single products. Lastly, the horizontal axis deals with the states and phases during the development of products.~\cite{bitkom2015strategy}

To enable the application of \ac{MBSE} for actually creating architectures of industrial systems based on \ac{RAMI}, the RAMI Toolbox has been proposed in~\cite{binder2021towards}. This framework is designed as an add-in for the modeling environment \ac{EA} and extends this software by a particular \ac{DSL} and an additional \ac{GUI}. The key goal of the RAMI Toolbox is to improve usability and support applicability. Thus, several functionalities are proposed. These functions include automatic model transformation, interfaces to other tools, and model-checking possibilities, among others. A key aspect is the reduction of the \ac{RAMI} 3D cube to a 2D grid, allowing for a more intuitive distinction of individual model views. The reduction is achieved by omitting the horizontal ``Life Cycle \& Value Stream'' axis and only considering the model valid for a specific phase of the life cycle.
\section{\ac{RAMI} model development and evaluation}\label{sec:imp}


In the following, we apply the \ac{IBPT} design approach to the Industry~4.0 scenario introduced in \cref{sec:scenario} and develop a \ac{RAMI} compatible model using \ac{EA}. We show that no potentially conflicting direct \ac{IT}/\ac{OT} interfaces exist in the resulting model, thus, according to Conway's law, eliminating potentially conflicting communication channels within the organization's communication structure.

The RAMI Toolbox proposed in~\cite{binder2021towards} is used as a tool to enable an \ac{MBSE} approach and ensure \ac{RAMI} compatibility as well as consistent traceability.
As the goal of the \ac{IBPT} integration at the system development phase is to eliminate potentially conflicting interfaces between \ac{IT} and \ac{OT} components, the \ac{DT} of the business process of playing the game of Nine Men's Morris provides interfaces abstracted to the level of game logic, substituting direct interfaces between \ac{IT} and \ac{OT} components. The proposed system architecture enables integration already during the design phase of the system, which allows early verification and validation. 
By following the principles of \ac{RAMI}, the architectures are aligned to the three dimensions of the reference model. The resulting \ac{EA} model is provided at~\cite{schaefer2023}.

\subsection{IBPT: Business processes integration}

The business process according to the scenario described in \cref{sec:scenario} is the creation of Nine Men's Morris game states. 
A game state consists of token positions on the game board and is established by game moves transmitted by humans or \ac{AI} agents.
We utilize \ac{OPCUA} to implement a holistic semantic information and communication model as a consequence to the interconnection, information transparency and decentralized decisions Industry~4.0 design principles outlined in \cite{hermann2016design}. \ac{OPCUA} offers a TCP/IP-based communication protocol as well as information modeling capabilities and is the preferred Industry~4.0 communication standard for the operating phase of production assets~\cite{heidel2019}.
Using these information modeling capabilities, we create interfaces that abstract interaction between devices to the level of game logic.
As a consequence, instead of transmitting target coordinates to robots, the communication is centered around the transmission of Nine Men's Morris game moves. In this particular context, the \ac{IT} components of the scenario encompass a web-based \ac{GUI} as well as \ac{NLP}, \ac{CV}, and \ac{AR} devices.
Furthermore, \ac{AI} agents can generate and transmit automatically generated game moves.
In the context of \cref{FIG_IBPT}, all these components interact with a central \ac{IBPT} entity.
Every interaction in the form of game moves triggers a process of establishing a desired game state.

The \ac{IBPT} element is used to decouple possible conflicting interfaces via abstraction to high-level business processes. In more detail, when considering the \ac{RAMI} Hierarchy axis levels and Interoperability layers, the enterprise architecture mainly comprises the Business layer along with the Connected World or Enterprise row. In contrast, if the Product row is omitted, the \ac{OT} architecture is placed within the bottom four layers of \ac{RAMI} and the bottom four rows of the automation pyramid Hierarchy levels. Thus, with regard to the \ac{IT} architecture, the \ac{IBPT} element abstracts those four layers, respectively rows, to the enterprise system. By digitizing these aspects and summarizing them into a single element, complexity is reduced within the entire system, which allows addressing enterprise stakeholders in their main area of knowledge.

\begin{figure}[tb]
\begin{center}
	\includegraphics[width=0.5\textwidth]{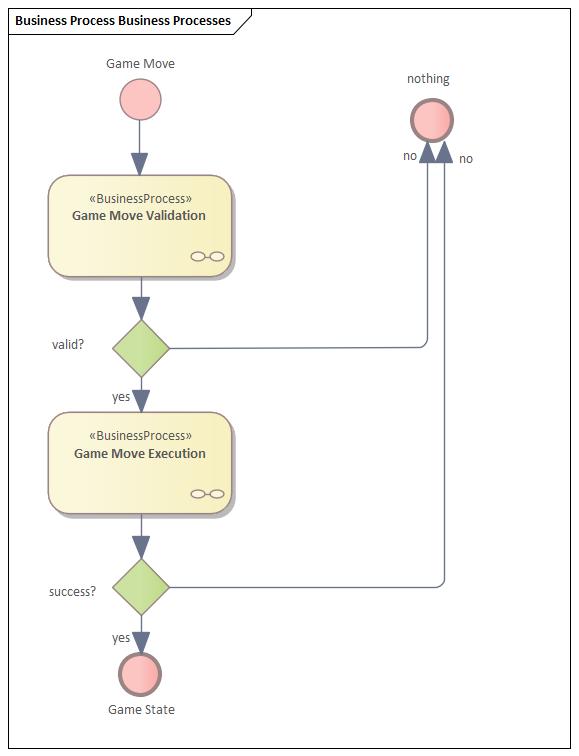}
\end{center}
	\caption{Nine Men's Morris Industry~4.0 scenario: Business process diagram.}
	\label{FIG_ITTEST}
\end{figure}

By doing so, the RAMI Toolbox proposes to utilize \ac{BPMN}-diagrams to represent the system's business processes and their sequences. Those processes directly address business cases or models of the system and indicate how the system should work or how stakeholders interact with it. Based on the business processes, requirements for developing and implementing the system are derived, which are also originating from stakeholder needs. Additionally, a system context model delimits the \ac{SoI} with boundaries and defines its in- and outputs. 
In this case study, we use \ac{BPMN} for implementing the \ac{IBPT} element. \cref{FIG_ITTEST} shows the business process of placing a Nine Men's Morris game move within the business process architecture of the scenario and the utilization of the \ac{IBPT} modeling element within this context. A game move production order is placed and validated at the \ac{IBPT} entity. If successful, the entity passes the game move to connected production units, controls information exchange and orchestrates the entire process flow.

\subsection{RAMI 4.0-based system architecture}

By creating the architecture before actually implementing the system, early verification and validation are enabled. In this way, we can investigate whether all potentially conflicting interfaces between \ac{IT} and \ac{OT} components have been eliminated before implementing the system.
When further modeling the reflection of the \ac{IBPT} element at the lower \ac{RAMI} layers, the physical implementation of the business process is described in detail. 
The robots along with their \acp{PLC} constitute the \ac{OT} architecture's core. The architecture allows for arbitrarily many robot cells (both physical and virtual) to be integrated, all replicating the current game state on their respective physical or virtual game boards. Regarding the Connected World or Enterprise row, \ac{SysML} is used for describing and implementing the \ac{IBPT} modeling element. In this case, it is necessary to embed the element into the bottom four layers of \ac{RAMI} and levels of the automation pyramid. While a \ac{DSL} is used to describe information or communication aspects of the system, the Integration and Asset layers are created by using \ac{SysML} block definition diagrams. These are also used when describing the hierarchical interrelation of the system components according to the automation pyramid.

\begin{figure}[tb]
\begin{center}
	\includegraphics[width=1\textwidth]{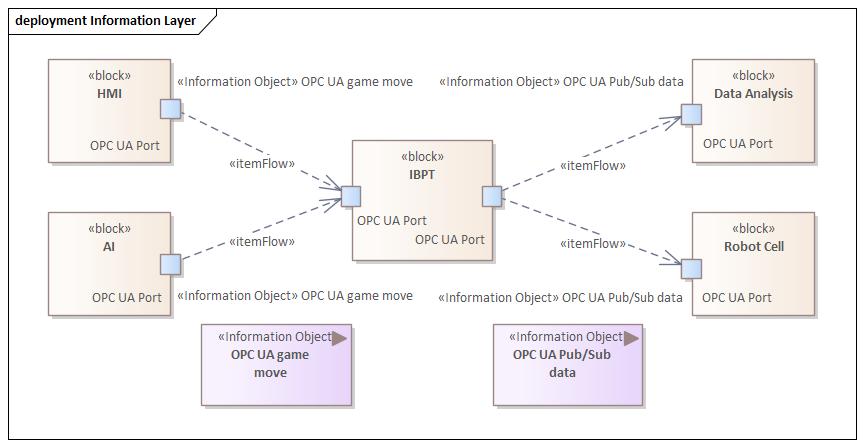}
\end{center}
	\caption{Nine Men's Morris Industry~4.0 scenario: Information layer diagram.}
	\label{FIG_INFLAY}
\end{figure}

\begin{figure}[tb]
\begin{center}
	\includegraphics[width=1\textwidth]{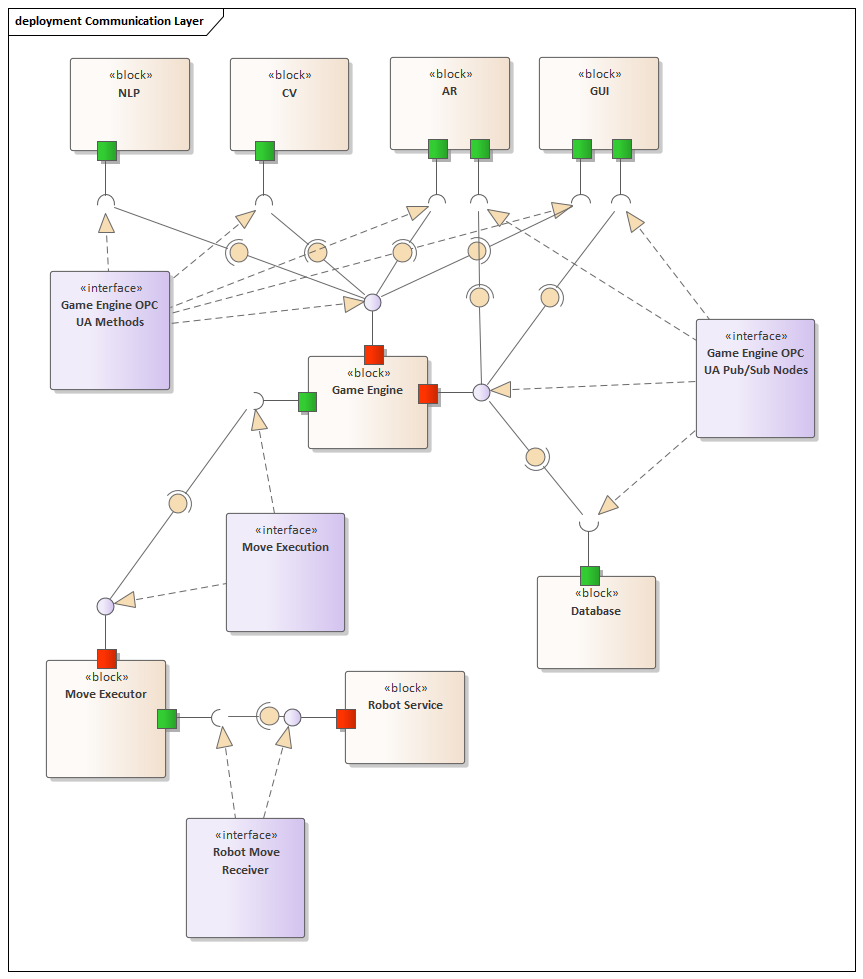}
\end{center}
	\caption{Nine Men's Morris Industry~4.0 scenario: Communication layer diagram.}
	\label{FIG_COMLAY}
\end{figure}

As far as the Information layer is concerned, the detailed information exchange between the components of the system is shown. 
\cref{FIG_INFLAY} indicates, how information flows are modeled. Game moves can be transmitted either via \acp{HMI} or \ac{AI} agents and are then transferred to the \ac{IBPT}, which deals with evaluating them and orchestrating connected production units accordingly. In parallel, created data is stored within a database for analysis purposes.
Since a \ac{SOA} is modeled, the Communication layer deals with provided and required services of the system, as shown in \cref{FIG_COMLAY}. A single \ac{OPCUA} interface, abstracted to the level of game logic, is utilized to place game moves at the \ac{IBPT}. Each of the \ac{HMI} and \ac{AI} components need to implement the required interface. The \ac{AR} as well as \ac{GUI} components subscribe to changes of a \ac{OPCUA} PubSub game state node provided by the \ac{IBPT}. In this way, the current game state can be retrieved and shown to the user. The database component subscribes to that node as well in order to log the game progress. 
Connected robots are represented by Robot Service components. Every Robot Service provides an \ac{OPCUA} interface in accordance with the \ac{OPCUA} information model defined by the \ac{IBPT}, for triggering token placement. Once a game move is validated, these interfaces are called by the \ac{IBPT}, which then tracks whether all robots execute the move correctly via the method call's return value.

\begin{figure}[tb]
\begin{center}
	\includegraphics[width=1\textwidth]{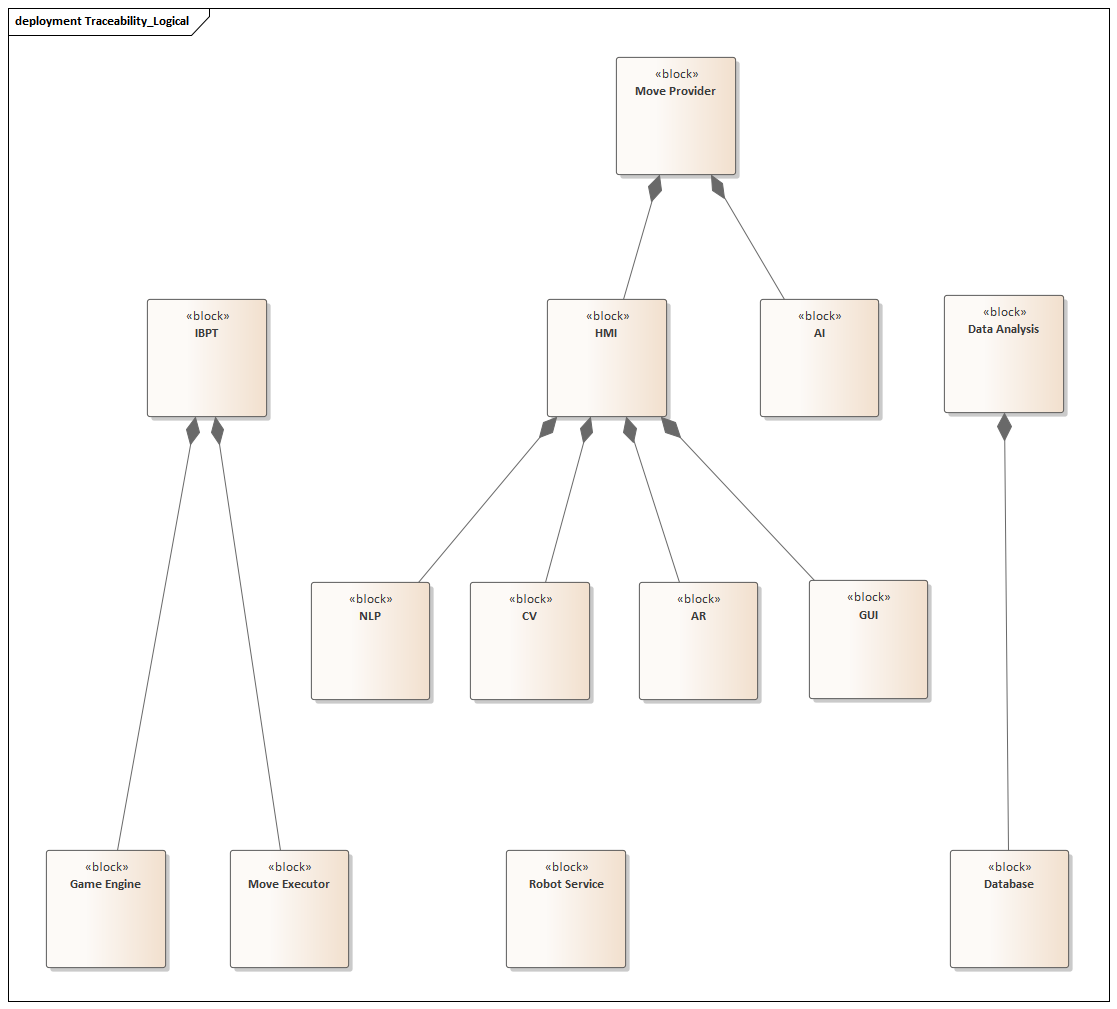}
\end{center}
	\caption{Nine Men's Morris Industry~4.0 scenario: Abstraction hierarchy diagram.}
	\label{FIG_ABSHIE}
\end{figure}

As far as the abstraction hierarchy is concerned, three different \ac{RAMI} levels are addressed: The Work Unit, Station and Control Device levels. At Work Unit level, game moves are provided by Move Provider components. As previously explained, this could be an \ac{AI} agent or various \acp{HMI}. The database is part of a Data Analysis component at station level. For this particular system design, the \ac{IBPT} element is constituted of a Game engine component, that handles all the game-related workflows, and a Move Executor component, in charge of executing game moves at Control Device level. The resulting hierarchy is shown in \cref{FIG_ABSHIE}.

Analyzing the resulting model of the system architecture provided at~\cite{schaefer2023} shows, that no direct interface connection between \ac{IT} and \ac{OT} component exist. This means that the entire logic is controlled via the business process / \ac{IBPT} and that potentially conflicting direct \ac{IT} and \ac{OT} interfaces have been eliminated by the \ac{IBPT} design approach.
\section{Conclusion and future work}\label{sec:conclusion}

By evaluating the \ac{RAMI} model of the presented Industry~4.0 scenario in \cref{sec:imp}, we show that the \ac{IBPT} approach eliminates potentially conflicting \ac{IT}/\ac{OT} interfaces in the system design. Assuming that Conway's law holds, this implies that conflicting communication channels also are reduced within the organization's communication structure. Although his theory has been widely accepted especially in the software development community, Conway, however, has never provided a formal or empirical proof of his theory. Although strong evidence supporting the theory was later published by MacCormack et al. in~\cite{macormack2012},
follow-up work on our paper could further provide evidence of this sociological phenomenon (and its inverse) via empirical studies conducted in production organizations, in order to further strengthen our argument.
Furthermore, outside the scope of \ac{IT}/\ac{OT} integration, future work could generalize our approach by specifically searching for conflicting communication channels within an organization and finding a system design that resolves these conflicts.

Evaluating the model of the described scenario shows that \ac{IT}/\ac{OT} integration is enabled by the \ac{IBPT} modeling element. Interaction between subsystems is abstracted to the level of business logic, which, in the case of our specific scenario, is playing the game of Nine Men's Morris. This makes interaction between subsystems easier to understand and handle, enabling a more intuitive understanding of the system for production process owners. The RAMI Toolbox enables a \ac{MBSE}-based approach, ensuring RAMI4.0 compatibility and consistent traceability. However, as \ac{MBSE} mainly addresses system architectures of instantiated systems during the design phase, the \ac{RAMI} \enquote{Life Cycle \& Value Stream} axis is out of context for this paper.
The integration of life-cycle aspects, mainly into \ac{OT}-oriented systems, needs to be investigated in the prospect of this work.

Future work could also investigate how the \ac{IBPT} entity could be realized as Industry~4.0
component compliant to the definition provided in~\cite{heidel2019}. This requires the extension of our data model to meet the requirements of \acp{AAS}. Specifications are provided by the \enquote{Plattform Industrie~4.0} association in~\cite{platformindustry402022_1} and~\cite{platformindustry402022_2} which define the base \ac{AAS} model as well as functional application programming interfaces.
%
%
%
\bibliographystyle{splncs04}
\bibliography{literature}
\end{document}